\documentstyle[11pt,ska,twoside,psfig]{article}
\begin{document}
\title{Pulsars with the SKA}
 \author{Michael Kramer}
\affil{University of Manchester, Jodrell Bank Observatory, \\
Macclesfield SK11 9DL, UK}

\begin{abstract}
As for other areas in modern astronomy, the SKA will revolutionize 
the field of pulsar astrophysics. Not only
will new science be possible by the shear number of pulsars discovered,
but also by the unique timing precision achievable with the SKA. The combination
of both will not simply mean a continuation of the successes already
achieved by using pulsars as fundamental tools of physics but the SKA
will provide a new quality of science.
\end{abstract}

\section{Introduction}

In the 35 years since the discovery of pulsars, these unique objects
have been proven to be invaluable in the study of a wide variety of
physical and astrophysical problems. Most notable are studies of
gravitational physics, the interior of neutron stars, the structure of
the Milky Way and stellar and binary evolution.  Most, but not all, of
these results have been obtained by pulsar timing. Other uses of
pulsars depend on measuring their emission properties and/or the
interaction of the radiation with an ambient medium. All such results
will be measurable with the SKA to unprecedented precision. But pulsar
astronomy will also benefit from the SKA simply because of the shear
number of sources that will be discovered and studied. Searching for
pulsars is a prime example where quantity results eventually in
quality by discovering a large number of exotic objects that probe
extreme physics. In the following I review the major aspects of the
science case for pulsars which is also outlined in the report of the {\em
Radio Transient, Stellar End Products, and SETI} working group. A more
detailed account of the opportunities provided by the SKA will be
presented in the forthcoming new SKA science case.

\section{Science with the SKA}

Studying the rotational behaviour and the propagation of the pulses in
curved space-time is achieved by {\em pulsar timing}. In this mode of
observing the arrival times of the pulses are measured with the
highest possible precision. The timing precision can be expressed by
the RMS of the residuals after fitting an appropriate spin-down model,
which scales as
\begin{equation}
\sigma \propto \frac{W}{\mbox{SNR}} \propto \frac{T_{sys}}{A_{eff}}
\times 
\frac{1}{\sqrt{2 \;\Delta \nu\; t }} 
\times \frac{W^{3/2}}{S_{psr}}
\end{equation}
since the signal-to-noise ratio scales as SNR $\propto 1/\sqrt{W}$,
roughly. The pulse width $W$ is measured in units of time, $T_{sys}$
is the system temperature on the sky, $A_{eff}$ the effective area of
the telescope, $\Delta \nu$ the available observing bandwidth, $t$ the
integration time, and $S_{psr}$ the flux density of the pulsar. The
last term in the above expression is source dependent, favouring
strong pulsars with narrow pulses (or narrow features that can be locked
on in a template matching procedure). Obviously, pulsars with a small
period (and hence $W$), so-called millisecond pulsars provide the highest
timing precision.

The special property of the SKA will be its unique sensitivity, i.e.~a
$T_{sys}/A_{eff}$ about 10 times smaller than for Arecibo or 100 times
smaller than for the Lovell telescope, Effelsberg or the GBT. Hence,
for the same sources it should in principle be possible to achieve a
timing accuracy with the SKA that is correspondingly better.  Apart
from the flux density of the pulsar to be timed, another limiting
factor will be the possible presence of timing noise. This random-like
variation in the pulsar's spin-down properties has been studied for
slowly-rotating pulsars, but is also now observed in some
faster-rotating millisecond pulsars
(e.g.~Lange et al.~2001, Lommen 2001). This suggests that every pulsar
may, when studied with sufficient precision, exhibit this
phenomenon. However, newly developed techniques may help to circumvent
this possible limitation (Hobbs et al.~2003),
and further studies will show how common it
will be observed for millisecond pulsars.

Before we can time pulsars, we have to find them in the first place.
The sensitivity of the SKA makes this a particularly exciting 
research area. The strawman-design of the SKA provides one with a
sensitivity of $S_{min}\approx 1.4\mu$Jy in only 1 min integration
time (assuming a detection threshold of $8\sigma$, a system temperature
of $T_{sys}\approx 25$K, and a bandwidth of $\Delta \nu \approx 0.5\nu$).
This corresponds to the luminosities listed in Table~\ref{tab:lum}.

\begin{table}[h!]
\caption{\label{tab:lum} 
Luminosities of pulsars detectable in a 1-min search with the SKA
at various distances. 
}
\medskip

\begin{center}
\begin{tabular}{lcc}
\hline
\hline
\noalign{\smallskip}
Where & D (kpc) & L (mJy kpc$^2$) \\
\noalign{\smallskip}
\hline
\noalign{\smallskip}
Galactic Centre & 8.5 & 0.1 \\
Opposite Side of Galaxy & 24 & 0.8 \\
Magellanic Clouds & 50--60 & 4.2 \\
M31 & 690 & 660 \\
\noalign{\smallskip}
\hline
\end{tabular}
\end{center}

\end{table}

At 1.4 GHz, the currently {\em observed} luminosities range from 0.01 mJy kpc$^2$
to 10,000 mJy kpc$^2$ with a (logarithmic) median of $\sim 25$ mJy
kpc$^2$.  Clearly, the luminosity limits achievable with the SKA 
at the low end will 
not only provide an essentially complete census of the Milky Way pulsars
but reasonable integration times even allow the discovery (and study!)
of pulsars in nearby galaxies. We will firstly consider the
implications resulting from the large number of pulsars that will
be discovered.

\subsection{Studying the pulsar population}

In seems possible that the largest part, if not all, of the Galactic
radio pulsars beaming towards Earth will be discovered with the
SKA, i.e.~about 20,000 pulsar could be found. 
This {\em Galactic Census} of pulsars opens up new horizons in
the study of a variety of subjects.

\subsubsection{Pulsar as neutron stars.}

The observations of relativistic effects in binary pulsars allows the
measurements of neutron stars masses.  Most straightforward is the
detection of a Shapiro time delay due to a companion for nearly
edge-on orbits. Currently, the number of binary systems with such
fortuitously aligned orbits is rather limited, but the combination of
number of binary pulsars discovered and the sensitivity with the
resulting timing precision will allow the measurement of many more
neutron star masses.  A manifold increase in the available statistics
will allow to study the amount of matter accreted during a spin-up
process and will also help in studying the neutron star
equation-of-state.
 
The study of the latter will also benefit from the vastly improved
statistics of observed pulse periods. Currently, the found periods
range from 1.5 ms to 8.5 s. The discovery of even smaller
rotational periods will provide significant insight in the properties
and stiffness of the ultra-dense liquid interior of neutron stars.
Even the case of a ``null result'', that no period shorter than 1.5 ms
will be found,  will have consequence as it requires the existence
of a limiting period not far away from the currently observed
value. At the other extreme end, the discovery of very long period
pulsars, isolated or in binary systems, will establish the connection
of radio pulsars to magnetars, Soft-Gamma ray Repeaters (SGRs) and
Anomalous X-ray Pulsars (AXPs), deciding as to whether these are
distinct classes of neutron stars or simply different evolutionary stages or
end products.

\subsubsection{Pulsars as radio sources.}

Despite (or perhaps, 'because of'!) many intensive studies of the
pulsar emission features occurring on timescales from nanoseconds to
hours, weeks and months, the nature of the emission process is still
unknown.  One of the problems is the need to separate intrinsic
effects from those caused by propagation. Moreover, it seems clear
that solving the pulsar emission mechanism requires the observations
of individual radio pulses. Only a limited number of pulsars is
currently known that are luminous enough for such experiments. In
particular, the needed multi-frequency observations are constrained by
the relatively small number of pulsars that can be studied with
sufficiently bright single pulses across a wide frequency range, and
by the limited availability of large sensitive telescopes. The SKA will
not only provide sufficient sensitivity at high frequencies
(depending on design), increasing the sample of pulsars
that can be studied to
hundreds or more, but a combination of sub-arrays and/or
multi-frequency feeds could perform these observations easily across a
wide range of frequencies.

Many more very faint pulsars will be detected and the luminosity
distribution of pulsars can finally be established. As we can expect
from the recent discovery of faint neutron stars in deep searches
(e.g.~Camilo et al.~2002), many more neutron stars observed at high
energies will have a detectable radio counterpart. Hence, the
simultaneous availability of new gamma-ray telescopes like GLAST will
mean that the relationship between the emission processes and 
plasma particles involved at the far ends of the electromagnetic
spectrum can be studied in great detail,

\subsection{Pulsars as probes}

\subsubsection{Mapping the Galaxy.}

The emission of radio pulsars interacts with the ionized magnetized
mediums it propagates through. The pulses become dispersed and
scattered while the position angle of the linearly polarized
emission component undergoes Faraday rotation depending on
electron density and magnetic field. The currently best model
for the electron density distribution in the Galaxy by Cordes \&
Lazio (2002) demonstrates impressively how information
about dispersion and scattering measures for a large number of
independent line-of-sights to pulsars located in different parts
of the Milky Way can be used to establish an increasingly detailed
knowledge. A Galactic Census of pulsars will provide so many 
lines-of-sight that, in particular with a combination of HI
absorption measurements, distances, electron densities and
scattering measures will be determined in such numbers
that very detailed
modelling becomes possible and a complete 3-D map of the Galaxy unfolds.

\subsubsection{Globular clusters.}

The  study of millisecond pulsars in globular clusters has
been rejuvenated in the last few years by a much increased number
of discovered pulsars, adding also to the list of clusters with
known pulsar content (e.g.~D'Amico et al.~2001). The example
of 47 Tucanae with more than 20 millisecond pulsars has provided
already a good example for the kind of science that becomes
possible by having the chance to study many more pulsars
in this or similar clusters (Freire et al.~2003): 
It will be possible to use these pulsars
to probe the clusters' gravitational potential, their evolutionary
stage, the stellar interaction and the intracluster medium. We
already achieve ``sub-milliarcsecond per year'' precision in
proper motion measurements and we start to see the relative motion
of the pulsars in the cluster. However, with the current
sensitivity and precision, further progress will be slow and it
will need another 10 years or so to break the 0.01 mas/yr boundary.
With an SKA we can achieve a proper motion measurement for 47TUC J,
the brightest MSP in 47Tuc, of sub-$\mu$as/yr precision in about 2 years
of weekly observations. This will also yield the distance of this
5-kpc cluster due to parallax measurements to better than 2\%.
This is a precision which even future astrometry missions like
GAIA will not be able to surpass.

\subsection{Gravitational Physics}

One of the most exciting motivations to find lots of new pulsars is the
prospect of finding some very exciting and exotic systems. The
available computer power will enable us to do much more sophisticated
searches than possible today, and the sensitivity allows much shorter
integration times, so that searches for fast accelerated pulsars
will not be limited anymore. Hence, the discovery rate for
relativistic binaries is certain to increase. But even in the worst
case scenario, with a detection rate remaining unchanged, we should
expect at least hundred relativistic binaries, or so, that provide
exciting test ground for gravitational physics.  We should find more
planetary systems, we may find (over-determined!) 
double pulsar systems and perhaps
even exotic or strange stars. Studying regions of high stellar density
with an increased probability of stellar interactions, such as globular
clusters and the Galactic centre (which is particularly interesting
due to its conditions favouring more massive stars) should provide us
with ``the holy grail'' of a pulsar-stellar black hole system, or
even with a millisecond pulsar orbiting a black hole. For the first
time it would be possible to probe the properties of a black hole
predicted by Einstein's theory of gravity.

\subsubsection{Properties of black holes}

The current best candidates for a stellar black holes are provided by
dynamical mass estimates in X-ray binaries. While it seems accepted
that any unseen companion with a mass exceeding $\sim3M_\odot$ is a
likely black hole candidate, the mass limit is somewhat uncertain due
to possible effects of rotation or even the speculated existence of
Boson- or even Q-stars for which no reliable mass estimate exists.
A realistic black hole should also rotate, which allows us 
for a black hole of mass $M$ to
define a (dimension less) spin, $\chi=cS/GM^2$, and quadrupole
moment, $q=c^4Q/G^2M^3$, whereas $S$ is the angular moment and
$Q$ is the quadrupole moment. In GR we have $\chi\le 1$ for
Kerr-black holes, while $\chi>1$ indicates a naked singularity.
In fact, a measurement of both $\chi$ and $q$ can be used to identify
the nature of the unseen companion, e.g.~for a Kerr-BH we find
$q\le -10\chi^2$, while for a neutron star, $q=-\chi^2$. A rotating
boson star may yield $q=-C\chi^2$ with $2.0\le C \le 12.1$
(Wex \& Kopeikin 1999).

It has been suggested that frame dragging could be measured by
pulsar timing to identify a black hole companion to a pulsar
in an edge-on orbit(Laguna \& Wolszczan 1997). 
Unfortunately this effect would be superposed on the usually
stronger light-bending effect (Doroshenko \& Kopeikin 1995)
which would make the detection of the frame dragging effect difficult.
Alternatively, the most promising way is to measure non-linear
changes in the longitude of periastron, $\omega$, and the projected
semi-major axis, $x$, caused by the gravitomagnetic field
(Wex \& Kopeikin 1999). Again,
the precessional effects to be searched for will be superposed
on other effects, but separating these effects on the way, will
provide an exciting tour de force through tests of gravity in
its own right. Competing effects for the observed change in $x$
are
\begin{equation}
\left( \frac{\dot{x}}{x} \right)^{obs} =
\left( \frac{\dot{x}}{x} \right)^{pre} +
\left( \frac{\dot{a}_p}{a_p} \right)^{gw} +
\left( \frac{\dot{x}}{x} \right)^{pm} +
\left( \frac{\dot{D}}{D} \right) +
\left( \frac{d\epsilon_a}{dt} \right)
\end{equation}
The first term is the one to be measured, the second term is due to
the shrinkage of the orbit due to gravitational wave emission, the
third term arises from a motion of the system relative to the observer
, the fourth term is due to an acceleration in an external
gravitational potential while the last term is due to geodetic
precession. All but the first term have been studied in pulsars and
binary systems already (e.g.~Weisberg \& Taylor 2003, van Straten et
al. 2001, Kramer 1998), so that with the increased precision of the
SKA we should be able to handle them even easier. Still, unless a
millisecond pulsar in orbit around a black hole can be found, the task
of measuring the black hole properties will even be a challenging 
for the SKA. But it is not impossible and it would be
surely among the most remarkable
discoveries that is likely to be made with the SKA.

\subsubsection{Cosmological Gravitational Wave Background}

Monitoring a large sample of millisecond pulsars in a so-called Pulsar
Timing Array (PTA) offers the means to detect the stochastic
gravitational wave background that is expected in various cosmological
theories (e.g.~Foster \& Backer 1990).  The PTA would be used to
search for (correlated) structures in the timing residuals of MSPs
distributed across the sky. It will be sensitive to long-wave
gravitational signals and is therefore complementary to current
ground-based gravitational wave detectors such as GEO600 or LIGO and
even future detectors such as LIGO-II or LISA. With a rough estimate
of about 1000 millisecond pulsars to be discovered, the SKA would not
only provide the necessary dense PTA but also the means of achieving
the necessary sub-$mu$s timing precision that will be needed to
extract the weak signal.

\begin{figure}[hbt]

\begin{tabular}{cc}
\psfig{figure=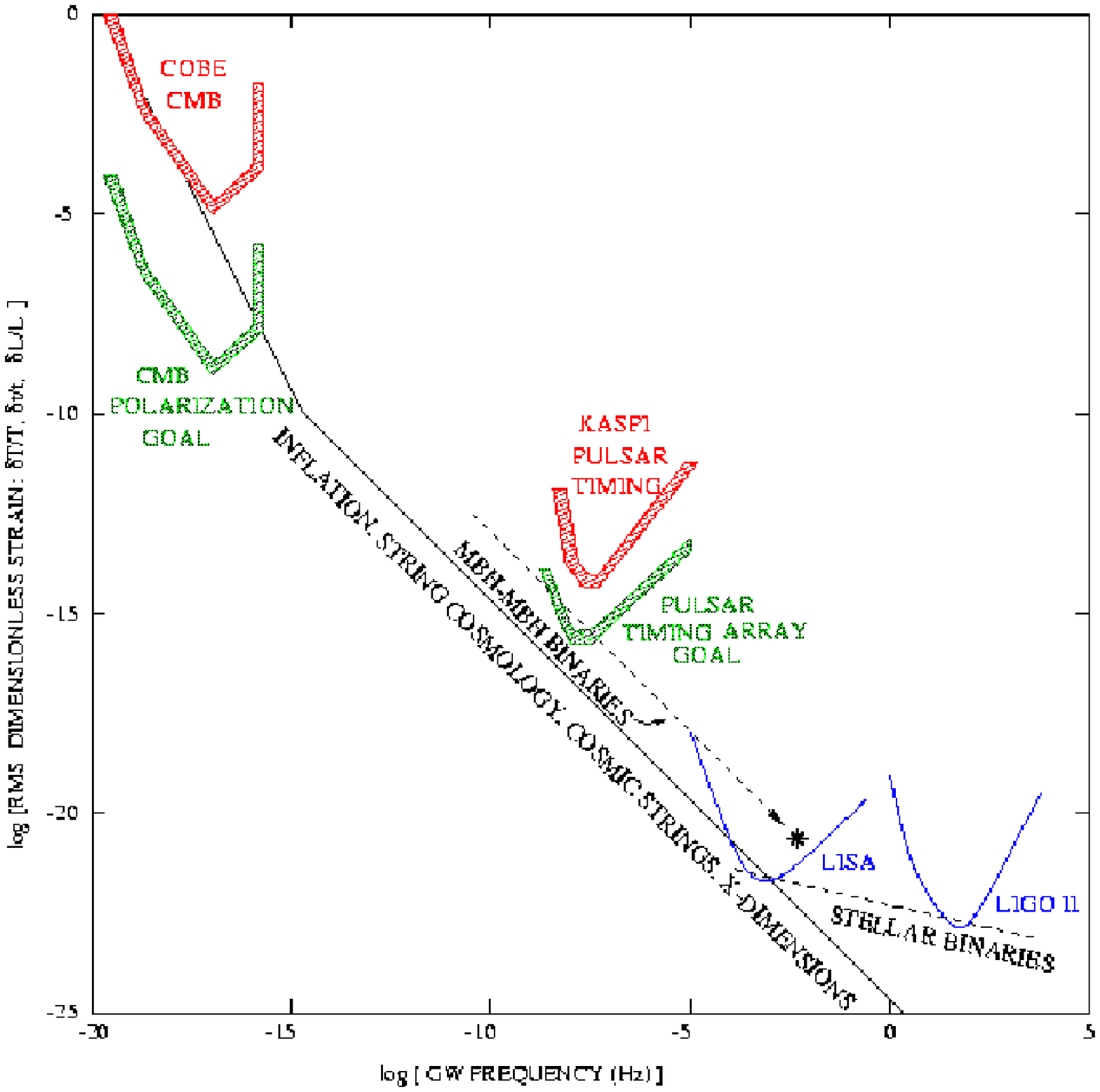,width=6cm} &
\psfig{figure=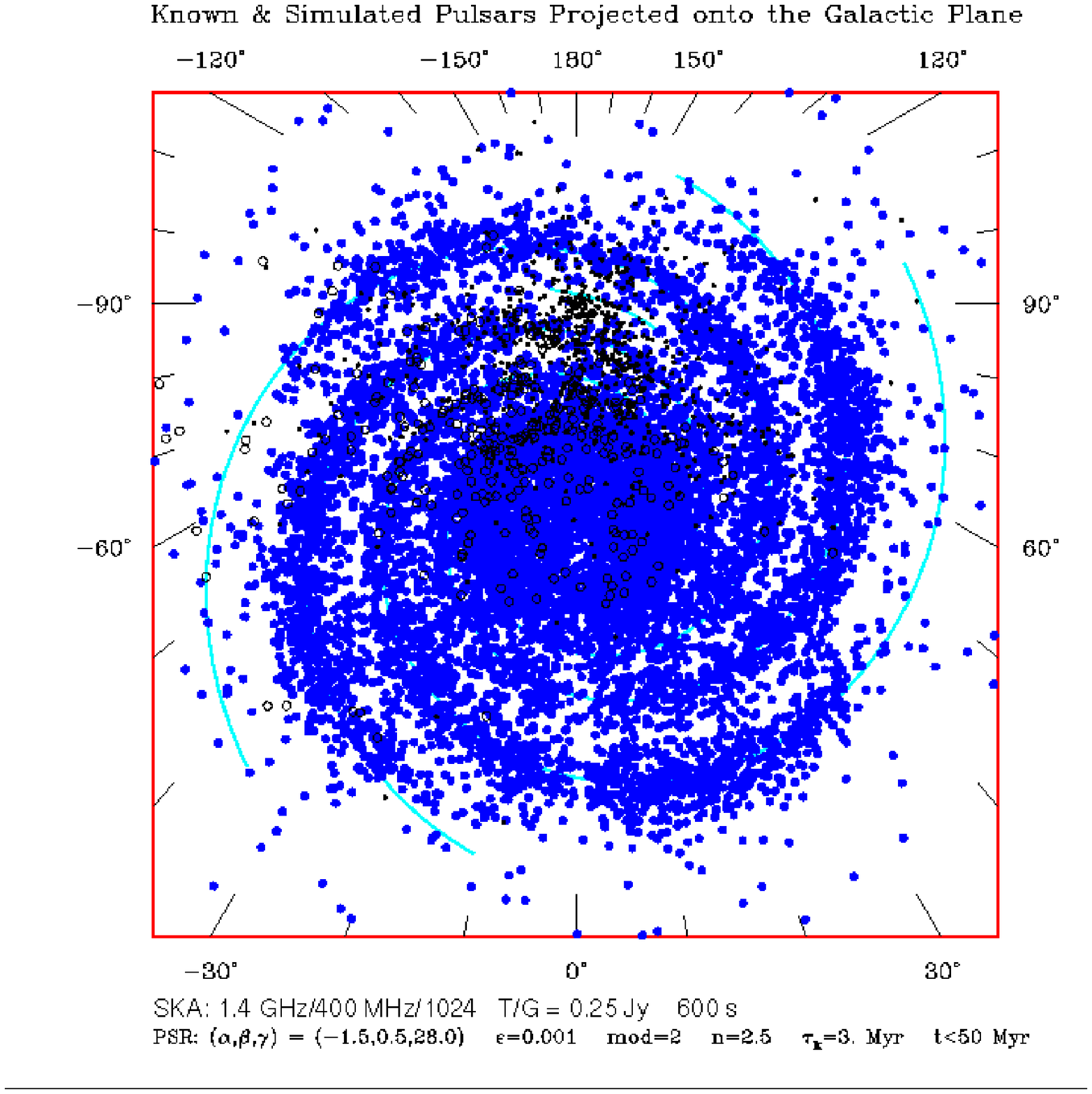,width=6cm} 
\end{tabular}

\caption{left) Sensitivity limits for detecting the gravitational
wave background (Backer, priv.~comm.), right) Galactic census
of pulsars discovered in a strawman SKA survey (Cordes, priv.~comm.).}
\end{figure}

\subsection{Extragalactic pulsars}

We have already seen above that standard searches with the SKA should
result in the discovery of pulsars in the most nearby galaxies,
allowing to study the population of pulsars as seen from the distance
and for different star formation histories. This search volume can be
further expanded if we search for giant pulses rather than for
periodic signals (e.g.~McLaughlin 2001). We could possibly
reach a distance of 5 Mpc or even 10 Mpc, 
and depending on the strength of the
pulses, larger distances are not impossible. Even though the study
(and timing!)  of these far away pulsars would be virtually
impossible, their giant pulses can still be used study the
distribution of pulsars in galaxies of different types, and to investigate
the ionized medium in the host galaxies
and intergalactic space.

\subsection{Value Added Science}

The above is clearly only a fraction of the science that will
be achievable with the SKA -- and that only in the area of pulsar
astrophysics! There is much more I should have added and which deserves
much more attention. Examples are the interior of neutron stars,
birth properties of pulsars,
neutron star-supernova associations, stellar, binary and
planetary evolution, core collapse physics, astrometry,
metrology, time keeping to name only a few.

\section{Conclusions}

Most of the science suggested here can be achieved with the proposed
strawman design. Simulations studying the multi-beaming requirements
for the pulsar case are underway, as the timing of about 20,000
pulsars is certainly a challenging task even for a multi-beaming
telescope.  Some long baseline capabilities are
desirable for astrometric purposes (1 mas at 5 GHz). 
The required upper frequency
limit is relatively modest but should be around 15 GHz which is necessary
to penetrate the scattering screen towards the inner Galaxy. 
Additionally, pulsar science sets tough requirements for the correlator
in terms of dump time (about $5\mu$s for timing, $50\mu$s for searching)
and bandwidth that can be processed.
Overall,
the science case in its potential and diversity appears overwhelming,
so that if there would not be a SKA anyway, there should be at least
one to study pulsars.

\acknowledgements
The science case is developed in collaboration with many colleagues
across the world, and in particular with those directly involved
in the Ratio Transient working group of the ISAC i.e.~Joe
Lazio (chair), Don Backer, Jim Cordes, Justin Jones, Michael Rupen,
and Jill Tarter.

\end{document}